\documentclass{moriond}

\def\be{\begin{equation}}
\def\ee{\end{equation}}
\def\bea{\begin{eqnarray}}
\def\eea{\end{eqnarray}}

\newcommand{\Photo}{}

\begin{document}
\title{Fast Radio Burst Cosmology and HIRAX}

\author{Amanda Weltman$^{1}$ and Anthony Walters$^{2}$}
\address{${}^{1}$Department of Mathematics and Applied Mathematics, University of Cape Town,  South Africa,\\${}^{2}$ Department of Chemistry and Physics, University of KwaZulu-Natal, Durban, South Africa }
\maketitle

\abstracts{Until very recently we had as many theories to explain Fast Radio Bursts as we have observations of them. An explosion of data is coming, if not here already, and thus it is an opportune time to understand how we can use FRBs for cosmology. The HIRAX experiment, based mostly in South Africa, will be one such experiment, designed not only to observe large numbers of FRBs but also to localise them. In this short article we consider briefly, some ways in which HIRAX can change the landscape of FRB cosmology.}

\section{Introduction}

Fast Radio Burst discovery is transitioning from a slow drip of data to a veritable deluge. In preparation for the expectation of a huge catalogue of FRBs and their dispersion measures, we consider the possible ways that they may be used as standard measures in cosmology as well as ways in which we can use FRB observations, cross correlated with cosmological surveys to answer other unsolved problems about our cosmos. In this short article based on a talk at the 2019 Moriond Gravity conference, we remind the reader about some of what is known about FRBs to date, as well as provide a brief introduction to the HIRAX experiment. We then go on to consider one such application for FRBs, the possibility of combining FRB results from HIRAX and Planck results to resolve the missing baryon problem.

\section{HIRAX}

The Hydrogen Intensity and Real-time Analysis experiment (HIRAX) is an array of 6m parabolic radio telescopes currently under construction in South Africa \cite{HIRAX}. The primary array will be situated in the Karoo desert area in South Africa, an area that has particularly low Radio Frequency Interference (RFI) due to an act of parliament, the Astronomy Advantages act of 2008 that protects the Karoo area from RFI from the usual manmade sources. The central HIRAX array will be based in the Karoo with outriggers planned most likely in other parts of the country - and in various other countries in Africa, most likely Rwanda and Namibia at the very least. HIRAX will be able detect the effects of dark energy on the distribution of galaxies by using Hydrogen intensity mapping in the 400 to 800 MHz range, which corresponds to a redshift range of  $0.8 < z < 2.5$. We expect that there are thousands, if not tens of thousands, of FRBs going off in some part of the observable sky every day. HIRAX will be able to observe a large fraction of these within its observing range, a 15 000 ${\rm deg}^2$ of the sky and will thus  produce a huge catalogue of several thousand FRBs. In its full design, HIRAX will be made up of 1024 dishes with several sets of outriggers, each made up of 8 dishes. An advantage of the Southern location of HIRAX allows for cross correlations with several ongoing surveys; see \cite{HIRAX} for further discussions on HIRAX setup and strengths. 

\section{Fast Radio Bursts}

Since their discovery a dozen years ago, Fast Radio Bursts have rapidly become an exciting novel area of astrophysics research, in part due to their relatively scarcity in the data so far and partly because we cannot explain their cause. If there is anything we love in astrophysics research, it is unexplained phenomena. In the case of Fast Radio Bursts, we have been using the sparse data we have so far to try to put together a picture of how they are formed, their expected rates and properties and whether they are all one type of object or fall in multiple classes. All of this is set to change in the coming months and years as globally we shift from searching for signals of FRBs in archival data, to observing them live, and in large numbers across the sky. At this stage, what we know is that they are most likely extragalactic and found within host galaxies. We observe them at relatively low redshifts up to about $z =0.5$ which is likely the a limitation of our ability to observe them. They are of the order of 1 Jansky in brightness and so far are very short lived, with lifespans ranging from $\mu {\rm s}$ to 50 ms and observed in the radio range as low as 400 MHz. There is a public catalogue of all bursts observed so far at {\tt www.frbcat.org} and a catalogue of theories at {\tt www.frbtheorycat.org} with a companion paper expanding on each so far \cite{FRBcatalogue}. They are observed at a range of polarisations, with none observed so far that are unpolarised, and they have a range of rotation measures. All of these clues do not yet point to a clear picture but we await more data. 

\section{Fast Radio Burst Cosmology}

Early indication that Fast Radio Bursts appeared to be extragalactic sources led to speculation on some possible cosmological applications of the bursts. And since the association of a repeating FRB with a host galaxy at $z=0.19$,  many more applications have been proposed. Even without redshift information some cosmological information an be extracted, for example; a single FRB can constrain violations of the Einstein Equivalence Principal \cite{2015PhRvL.115z1101W,tingay},  or constrain the mass of the photon \cite{2016ApJ...822L..15W}. Strongly lensed FRBs could be used to probe dark matter \cite{2016PhRvL.117i1301M}, or measure the value Hubble constant and cosmic curvature \cite{2017arXiv170806357L}. And, dispersion space distortions could be used to probe the clustering of matter \cite{2015PhRvL.115l1301M}.  In the future, should more FRB events be associated with a host galaxies, for which redshifts can be acquired,  this would give access to the Dispersion Measure (DM) redshift relation, $\mathrm{DM}(z)$, which can be used as a probe of the background cosmological parameters \cite{2014PhRvD..89j7303Z,2014ApJ...783L..35D,2014ApJ...788..189G}. However, the strength of this approach will strongly depend on the intrinsic scatter in the observed DM data caused by intervening matter inhomogeneities along the line of sight,  host galaxy contamination to the observed DM, and knowledge of the cosmic mean gas fraction in the IGM \cite{ourFRBpaper}.  Some other cosmological  applicants include constraining the growth rate by cross-correlating FRBs with kSZ data \cite{newJonPaper}, and testing the Copernican principal by testing the isotropy in the $\mathrm{DM}(z)$ relation \cite{newIsoPaper}.

\section{The Missing Baryon Problem}

Despite our current era of concordance cosmology, a number of puzzles remain. Included in these is the missing baryon problem \cite{Fukugita}; despite our best efforts and most sophisticated analyses, it appears that we have lost 30\% of the baryons we expect to see in the recent universe \cite{2004ApJ...616..643F,2012ApJ...759...23S}. While we believe they are likely in the warm-hot intergalactic medium (WHIM), gaining direct observational evidence of this is challenging.  Attempts at detecting the WHIM include large-scale structure cross-correlations, such as the cross-correlation between thermal Sunyaev-Zeldovich effect and galaxy weak lensing~\cite{2015JCAP...09..046M,Hojjati15,Hojjati17}, stacking  luminous red galaxy pairs with thermal SZ map~\cite{Tanimura19}, detecting the temperature dispersion of kinetic SZ effect within the X-ray selected galaxy clusters~\cite{Planck-dispersion2018}, and the detection of the cross-correlation between kinetic SZ effect with velocity field~\cite{Planck-unbound16,Carlos15}.  Since an FRBs DM is due to its propagation through regions with free electrons, they are directly sensitive to the location of baryons in the Universe, and thus may help to constrain the mass fraction of the WHIM. One such proposal considers cross-correlating FRB maps with the thermal Sunyaev-Zeldovich effect to find missing baryons \cite{2018PhRvD..98j3518M}.  Another approach would be to use the $\mathrm{DM}(z)$ relation to constrain the mean cosmic gas (diffuse baryons) in the IGM, $f_\mathrm{IGM}$ \cite{newFRBpaper}.

\section{Future Constraints from FRB observations}
\begin{figure}[h]
 \centering
    \includegraphics[width=.65\textwidth]{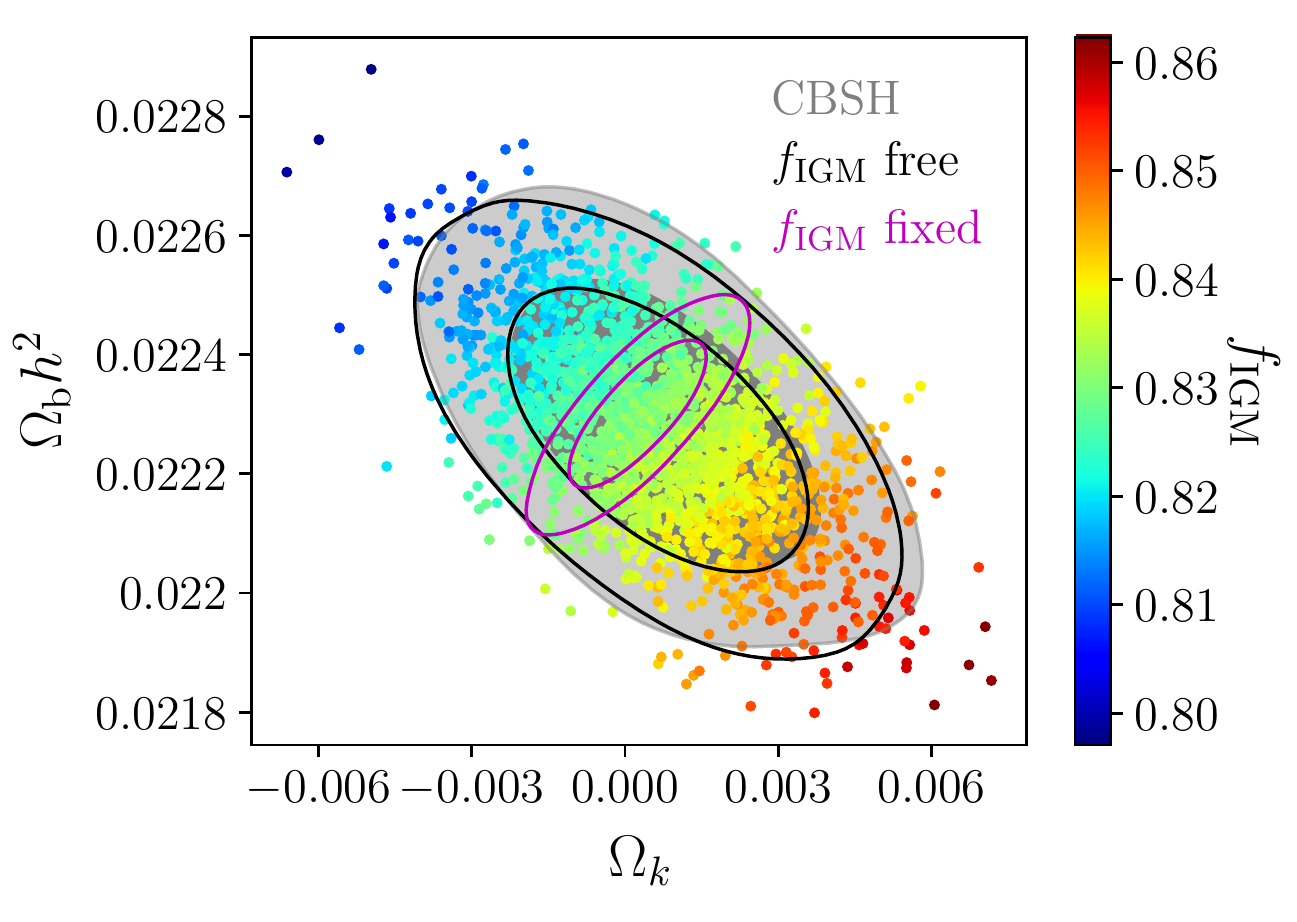}
  \caption{Constraint forecast in the $\Omega_k-\Omega_\mathrm{b}h^2$ plane, from $10^4$ FRBs with redshifts, using the DM(z) relation and Planck priors, and.... Grey contours show the CMB + BAO + SNe + $H_0$ (CBSH) constraint from the Planck 2015 results. Magenta lines indicate the 1- and 2-$\sigma$ constraint contours for FRB+CBSH assuming one has perfect knowledge of the mean diffuse baryon fraction in the IGM, $f_\mathrm{IGM}$. Black lines show the same, but with no prior knowledge of $f_\mathrm{IGM}$. Coloured points correspond to the value of $f_\mathrm{IGM}$. }
  \label{fig:ellipses}
 \end{figure}

A forecast using the $\mathrm{DM}(z)$ relation can be seen in Figure \ref{fig:ellipses}, which shows constraints  in the $\Omega_k-\Omega_\mathrm{b}h^2$ plane, from a sample of $10^4$ simulated  FRBs with redshifts, combined with the Planck 2015 constraints \cite{newFRBpaper}. Grey contours show the CMB + BAO + SNe + $H_0$ (CBSH) constraint from the Planck 2015 results. Magenta lines indicate the 1- and 2-$\sigma$ constraint contours for FRB+CBSH assuming one has perfect knowledge of the mean diffuse baryon fraction in the IGM, $f_\mathrm{IGM}$.   Clearly, having perfect knowledge of $f_\mathrm{IGM}$ would allow for a dramatic improvement over the current CBSH constraints, however this function in poorly constrained by observations. Black lines show the same, but with no prior knowledge of $f_\mathrm{IGM}$. While the cosmological constraints do not shown any improvement over the CBSH priors,  the priors allow for a measurement of $f_\mathrm{IGM}$ at the percent level (shown by the colorbar on the right). Since around $50-80\%$ of the baryons are believed to reside in the IGM, a detection of $f_\mathrm{IGM}$ in this range would provide further evidence that the most of the Universe's baryons are located in diffuse gas in the IGM, and are in-fact not missing. 

\section{Conclusion} From only a few FRBs observed in the last few years, we expect to soon have a catalogue of tens of thousands (if not far more) FRBs from radio telescope arrays around the world, including CHIME, ASKAP, and HIRAX as we have considered here. In this short article we have considered some novel cosmology and astrophysics questions we can probe using the incoming barrage of FRB data in the near future. We have highlighted the potential for localising these FRBs in particular with HIRAX, due to the comprehensive outrigger programme planned.

\section*{Acknowledgements}

This work is based in part on a talk given by Weltman at the 2019 Moriond Gravity meeting, for which she would like to thank the Organisers. We gratefully acknowledge our collaborators on the projects referenced in this work, \cite{Amadeuspaper} and \cite{ourFRBpaper}, and \cite{newFRBpaper}. Weltman is supported by the South African Research Chairs Initiative of the Department of Science and Technology and the National Research Foundation of South Africa (NRF). Any opinion, finding and conclusion or recommendation expressed in this material is that of the authors and the NRF does not accept any liability in this regard. Walters acknowledges support from the NRF (Grant Number 105925,  110984, 109577).  

\end{document}